\documentclass{aa}
\usepackage{txfonts}
\usepackage{natbib}
\usepackage{graphicx}

\newcommand{\Fe}{\mbox{\ion{Fe}{ii}}}
\newcommand{\Ox}{\mbox{[\ion{O}{iii}]}}
\newcommand{\Hb}{\mbox{H$\beta$}}
\bibpunct{(}{)}{;}{a}{}{,}

\begin{document}
\title{Extended emission-line regions in low-redshift quasars: Dependence on 
       nuclear spectral properties%
\thanks{Based on observations collected at the Centro Astron\'omico
Hispano Alem\'an (CAHA) at Calar Alto, operated jointly by the Max-
Planck-Institut f\"ur Astronomie and the Instituto de Astrof\'isica de
Andaluc\'ia (CSIC)}}

\author{Bernd Husemann\inst{1}
   \and Lutz Wisotzki\inst{1} 
   \and Sebastian F. S\'anchez\inst{2} 
   \and Knud Jahnke\inst{3}
}

\offprints{bhusemann@aip.de}
\institute{Astrophysikalisches Institut Potsdam, An der Sternwarte 16, 14482 Potsdam, Germany  
\and
Centro Astron\'omico Hispano Alem\'an de Calar Alto (CSIC-MPIA), E-4004 Almer\'ia, Spain
\and 
Max-Planck-Institut f\"ur Astronomie, K\"onigsstuhl 17, D-69117 Heidelberg, Germany}

\date{}
\abstract{}
%Aims
{We searched for the presence of extended emission-line regions (EELRs) around
  low-redshift QSOs.}
%Method
{We observed a sample of 20 mainly radio-quiet low-redshift quasars ($z<0.3)$
  by means of integral field spectroscopy. After decomposing the extended and
  nuclear emission components, we constructed \Ox\ $\lambda5007$ narrow-band
  images of the EELR to measure the total flux. From the same data we obtained
  high S/N ($>$50) nuclear spectra to measure properties such as \Ox/\Hb\ flux
  ratios, \Fe\ equivalent widths and H$\beta$ line widths.}
%Results
{A significant fraction of the quasars (8/20) show a luminous EELR, with
  detected linear sizes of several kpc. Whether or not a QSO has a luminous
  EELR is strongly related with nuclear properties, in the sense that an EELR
  was detected in objects with low \Fe\ equivalent width and large H$\beta$
  FWHM.  The EELRs were detected preferentially in QSOs with larger black hole
  masses. There is no discernible relation, however, between EELR detection
  and QSO luminosity and Eddington ratio.}  {}{}


\maketitle

\section{Introduction}

Quasars may show extended emission-line regions (EELRs) stretching
over characteristic scales of ten to hundred kpc. Luminous EELRs 
have been found in particular around steep-spectrum radio-loud quasars
\citep[RLQs,][]{Boroson:1984,Heckman:1991,Crawford:2000,Stockton:2002,Fu:2006}
and also radio galaxies
\citep[e.g.][]{Villar-Martin:2006,Villar-Martin:2007}.
There are also indications that radio-quiet quasars (RQQs) can have EELRs
\citep[e.g.,][]{Boroson:1985,Stockton:1987,Bennert:2002}. Most existing
studies, however, are based either on long-slit spectroscopy or on 
narrow-band imaging; while the former technique will generally capture 
only part of any extended emission, the latter does not provide any
spectroscopic information.

Integral field spectroscopy is a relatively new and powerful tool
to study AGN host galaxies. The combined diagnostic power of imaging and
spectroscopy allows for a flexible treatmeant of the data, facilitating the
construction of broad- or narrow-band images, of 2-dimensional velocity maps
as traced by emission or absorption lines, or the spectral analysis of
selected regions within the field of view.

Here we present preliminary results from a study of 20 low-redshift QSOs, most
of which are RQQs, observed with an integral field unit (IFU).  After
decomposing nuclear and extended emission we identified 8 quasars with clearly
detectable EELRs. We focus on exploring the relation between the nuclear
spectral properties of a QSO and the existence or non-existence of a
luminous EELR. A full account of our IFU study will be given in a separate
paper (Husemann et al., in preparation; hereafter Paper~II).

\section{Observations and Data Reduction}

We targeted a sample of 20 low-redshift ($z<0.3$) QSOs, selected from the
Palomar-Green Survey \citep[PG,][]{Schmidt:1983} and from the Hamburg/ESO
Survey (HES; \citet{Wisotzki:2000}; Wisotzki et al., in prep.). The QSOs
have apparent magnitudes in the range of $V\sim 14-17$; corresponding to
typical luminosities of $M_V < -23$.  Three objects are steep-spectrum RLQs,
another one is classified as a flat-spectrum RLQ; all the other sources are
known to be radio-quiet.  To our knowledge, this is the largest sample of
bright quasars observed with an IFU so far.

We employed the Potsdam Multi-Aperture Spectro\-photometer
\citep[PMAS;][]{Roth:2005}, mounted on the 3.5~m telescope at Calar Alto 
Observatory, for two observing runs in September 2002 and in May 2003.
The PMAS lens array was used at a spatial pixel (`spaxel') scale of 
$0.5\arcsec \times 0.5\arcsec$, which for $16\times 16$ spaxels
resulted in a field of view of $8\arcsec \times 8\arcsec$. At the
typical redshifts of our targets we were able to sample the
quasar environment with a resolution of $\sim 1$~kpc, and cover 
linear extents of up to $\sim 20$\,kpc in diameter.

All observations used the V300 grating to map
the spectral features around the H$\alpha$ and H$\beta$ emission lines in a
single setup. This yielded a moderate spectral resolution of
approximately 6.5\,\AA\ FWHM as measured by night sky emission lines.  The
total integration time for each object was 1--2 hours divided into several
30-45\,min. exposures.  Each individual exposure reaches a S/N level for the
integrated nuclear quasar spectrum of already more than 50 per pixel in the
continuum.

The raw data were homogeneously reduced using the PMAS reduction package
\textit{P3D} by \citet{Becker:2002}. Due to substantial fibre-to-fibre
variations in the wavelength calibration of more than 0.7~\AA\ rms, we used
elements of the general reduction package \textit{R3D} for fibre-fed IFUs
\citep{Sanchez:2006a} to reduce this scatter to 0.25~\AA\ rms.  We extracted 
and subsequently subtracted a mean sky spectrum from those spaxels in each 
science datacube that were deemed free of emission from the target itself. 
The data reduction will be described in full detail in Paper~II.

\section{Results}

\subsection{Quasar-EELR decomposition}

In order to measure the properties of EELRs, any possible contamination by the
quasar nuclear emission had to be minimised. Three components are
expected that need to be carefully deblended: Nuclear emission from the AGN
point source; the stellar component of the host galaxy; and any possible
resolved extended line emission. We found that in our luminous quasars, no
significant evidence for a resolved stellar continuum above the noise could be
found in our data, and we only needed to deblend the nuclear source
and the extended line emission.

After running several tests we adopted the iterative decomposition method by
\citet{Christensen:2006}. Briefly, a quasar spectrum is extracted from the
spaxels at the position of the quasar, assuming that the contribution of the
EELR and the stellar component is negligible for these spaxels. By dividing
the spectrum of each spaxel by the quasar spectrum in the wavelength range
bracketing a broad emission line (e.g. H$\alpha$ or H$\beta$), an appropriate
scale factor is determined for each spaxel.  A complete datacube of the quasar
emission using the previously determined scale factors is then constructed. By
subtracting the quasar datacube from the original one, a residual datacube
containing any possible EELR is obtained. The process can be iterated 
by extracting a new quasar spectrum from the previous reconstracted quasar datacube; however,
our solution converged already after the first iteration.
This method showed the best results
and minimal PSF subtraction artefacts compared to other strategies such as
off-band image subtraction or PSF synthesis \citep{Jahnke:2004,Sanchez:2004a}
that we also explored.  However, the method is not perfect, and there are
frequently some oversubtraction signatures at the quasar position (cf.\
Fig.~\ref{fig:images}). This happens because the extracted quasar spectrum contains
a small fraction of the extended narrow emission line even if the scale factors were
properly determined by the broad H$\beta$ line.

For the present paper we focus on the narrow wavelength range containing the
H$\beta$ and the \Ox\ doublet $\lambda\lambda 4959,5007$. The scale factor can
therefore safely assumed to be a constant for each spectrum. Collapsing the
residual data cube within $\pm 20$~\AA\ centred on the nuclear \Ox\ peak
wavelength, we could then produce narrow band \Ox\ images for each quasar.
Figure~\ref{fig:images} presents the results, displaying all 8 quasars with a
detected EELR in \Ox. In the bottom row we also show the residuals for a
representative subset of 3 (out of 12) undetected sources, to give the reader
a visual impression of the significance of the detections.

\subsection{EELR fluxes}

\begin{figure}
\resizebox{\hsize}{!}{	
  \includegraphics[bb=139 348 403 704]{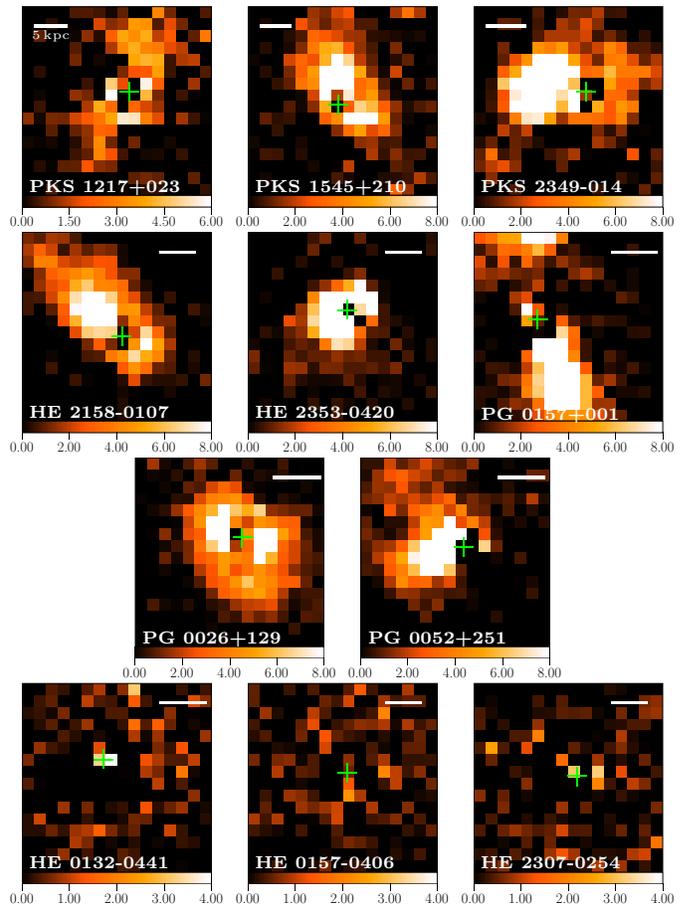}
  }
\caption{Nucleus-subtracted \Ox\ narrow-band images extracted from the
  datacubes.  All 8 quasars with clearly detected EELR are shown in the top
  three rows. The bottom row presents three example cases with non-detected
  EELR for comparison. In each panel, the position of the quasar is indicated
  by a green cross. Note that the apparent holes at most quasar locations are
  due to oversubtraction and thus not real (see text for details).  The colour
  coded line flux is in units of
  $10^{-16}\,\mathrm{erg/s/cm}^2\mathrm{/arcsec}^2$. All frames are 8\arcsec\
  on the side and oriented such that N is right and E is up.  The white marker
  shows a 5kpc linear scale at the quasar redshift.}
\label{fig:images}
 \end{figure}

Examining the extracted \Ox\ narrow band images after subtraction of the 
quasar nuclei, we found that a significant fraction (8/20) of our objects 
contain a luminous EELR. 
We measured the total fluxes in these extended emission components by
summation of all spaxels in the narrow band images, except those used to
extract the quasar spectrum as these spaxels are slightly affected by
oversubtraction of the quasar emission.  In the case of \object{PG~0157+001},
the EELR is evidently larger than the PMAS field of view (see
Fig.~\ref{fig:images}), and a significant fraction of the total flux may be
missed by our data. We therefore treat the measured EELR flux for this object
as a \emph{lower limit}. In all other cases we find the EELRs to be reasonably
well contained within the recorded images.

For the objects showing no sign of an EELR we could only estimate upper limits
to fluxes. We simply adopted the extended flux from the
faintest EELR measured in our sample as the upper limit for all those objects
where there was no EELR detected. This is clearly a very conservative
assumption which we plan to improve in the future by running dedicated
simulations. The adopted limit was then scaled to the continuum flux density
at 5100\,\AA\ of each quasar continuum to account for the contrast between
nuclear and extended emission. 

Three of our QSOs were already imaged through narrow-band filters by
\citet{Stockton:1987}. Despite the different observing techniques, our \Ox\
EELR luminosities agree within a factor of $\sim 2$--3 with their
measurements. Three QSOs were furthermore contained in the narrow-band HST
study by \citet{Bennert:2002}, which however revealed mainly high-surface
brightness features less than 1\arcsec\ to the QSO nuclei.

In order to obtain a measure of EELR strength independent of the
spectro-photometric calibration, Galactic foreground extinction and 
scaling with quasar luminosity, we divided the EELR fluxes (and upper 
limits) by the fluxes in the broad H$\beta$ components of the 
individual quasars.

\subsection{Spectral properties of the quasars}

As mentioned above, each datacube provided us also with a high S/N spectrum of
the central quasar over essentially the full optical wavelength range, albeit
at moderate spectral resolution. We extracted the nuclear emission line
properties by fitting a superposition of simple analytical functions to the
line profiles, using the Levenberg-Marquardt $\chi^2$ minimisation algorithm.
For our narrow wavelength range, a straight line was
sufficient to model the local continuum of each spectrum. The broad Balmer
lines were modelled with two or three independent Gaussian components, in
accordance with previous experiences by several others
\citep[e.g.][]{Sulentic:2002,Veron-Cetty:2006}.  We fitted the \Ox\
$\lambda\lambda$4959, 5007 doublet as a system of two Gaussians with the same
line width, fixed rest-frame wavelength separation and a fixed line ratio of 1:3.

Both H$\beta$ and the \Ox\ doublet can be heavily affected by \Fe\ emission
line complexes. The most prominent is the \Fe\ $\lambda\lambda
4924,5018$ doublet, which we modelled similarly to the \Ox\ doublet, but with
a fixed line ratio of 1.28 ($\lambda5018/\lambda4924$). We also used the \Fe\ complex 
($5100$\AA--$5405$\AA) longwards of \Ox\ to measure the strength of 
the \Fe\ emission in those cases where the \Fe\ emission is weak. 
A linear quasi-continuum was estimated from the bracketing emission-line free 
regions and subtracted from the quasar spectrum. We integrated the 
residual flux in the given wavelength interval to estimate the \Fe\ flux and 
equivalent widths.

\begin{figure}
  \centering
   \resizebox{\hsize}{!}{
   \includegraphics[height=4.5cm]{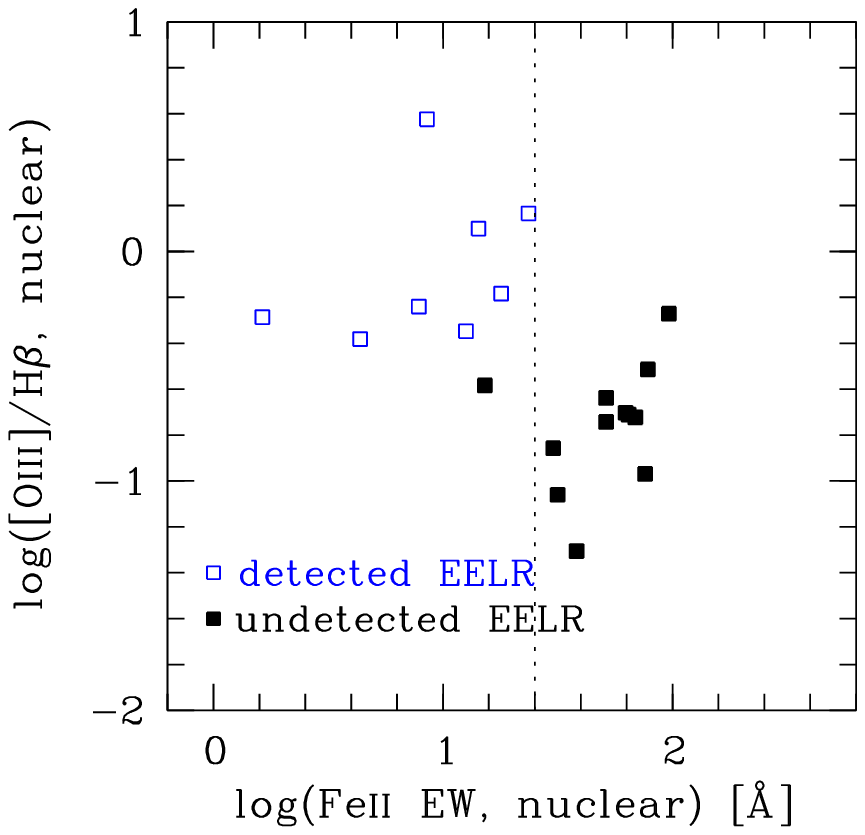}
   \includegraphics[height=4.5cm]{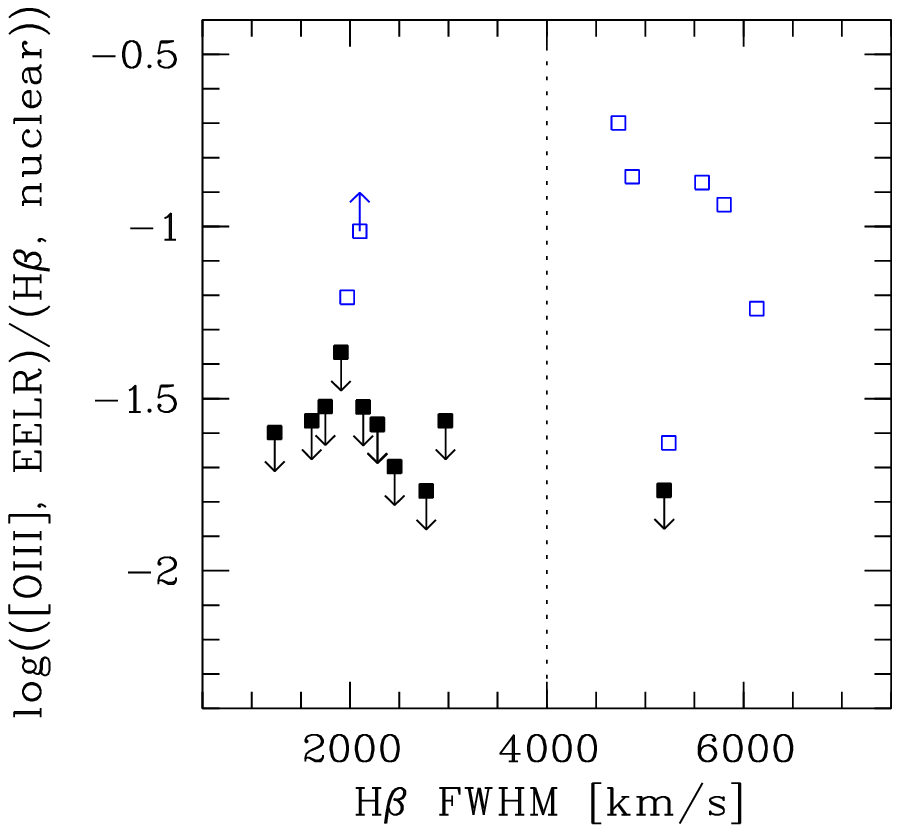}
   }
  \caption{Left panel: Nuclear \Ox/\Hb\ flux ratio plotted against 
      the nuclear \Fe\ equivalent width. QSOs with a detected EELR are 
      denoted by the open squares, the undetected ones are shown as upper
      limits. The dashed line indicates a suggestive separation between the 
      detected and non-detected EELRs. 
      Right panel: Total EELR flux, normalised to the nuclear \Hb\ flux,
      plotted against the FWHM of the broad H$\beta$ component.
      Note the \emph{lower} limit for PG~0157$+$001 for which  
      the EELR is larger than our field of view.}
  \label{fig:blr}
\end{figure}

\subsection{Black hole masses and accretion rates} \label{sec:bhmasses}

We estimated black hole masses ($M_\mathrm{BH}$) for all objects from our
single epoch quasar spectra, using the virial method
\citep{Vestergaard:2006}. The method assumes that the broad line region (BLR)
clouds are in gravitational equilibrium within the black hole potential, so
that the central mass can be estimated as $M_\mathrm{BH} = fRv^2/G$ where $f$
is a geometry-dependent factor that needs to be calibrated empirically.  The
radius $R$ of the BLR can be predicted from the quasar luminosity using the
radius-luminosity scaling relation $R \propto L^{0.5}$ \citep{Bentz:2006}.
The characteristic velocity of the emitting clouds can be estimated in two
ways, from the FWHM of broad emission lines or from the emission line
dispersion $\sigma$. \citet{Collin:2006} showed recently that the
$M_\mathrm{BH}$ estimates are sensitive to the line profile shape and argued
that $\sigma_\mathrm{line}$ gives more robust $M_\mathrm{BH}$ estimates than
those based on $\mathrm{FWHM}_\mathrm{line}$. We compared both velocity
measures and found that while the differences were non-negligible, they are
not important for the purpose of the present paper. We therefore adopted the
H$\beta$ line FWHM together with the scaling relation by
\citet{Vestergaard:2006} to obtain black hole masses.

Having obtained black hole masses, we could also estimate the `Eddington
ratios' of each quasar, i.e.\ the actual accretion rates divided by the
Eddington rates, or in terms of observables, the bolometric luminosities
divided by the luminosities that each object would have if radiating
at the Eddington limit, $L_\mathrm{bol}/L_\mathrm{Edd}$. 
We adopted a canonic bolometric correction \citep[e.g.,][]{Kaspi:2000}
of $L_\mathrm{bol}=9\lambda L_\lambda(5100$\AA).

\section{Discussion}

\subsection{Relations between EELR detection and nuclear spectral properties}
\begin{figure}
  \centering
   \resizebox{\hsize}{!}{
   \includegraphics[height=4.5cm]{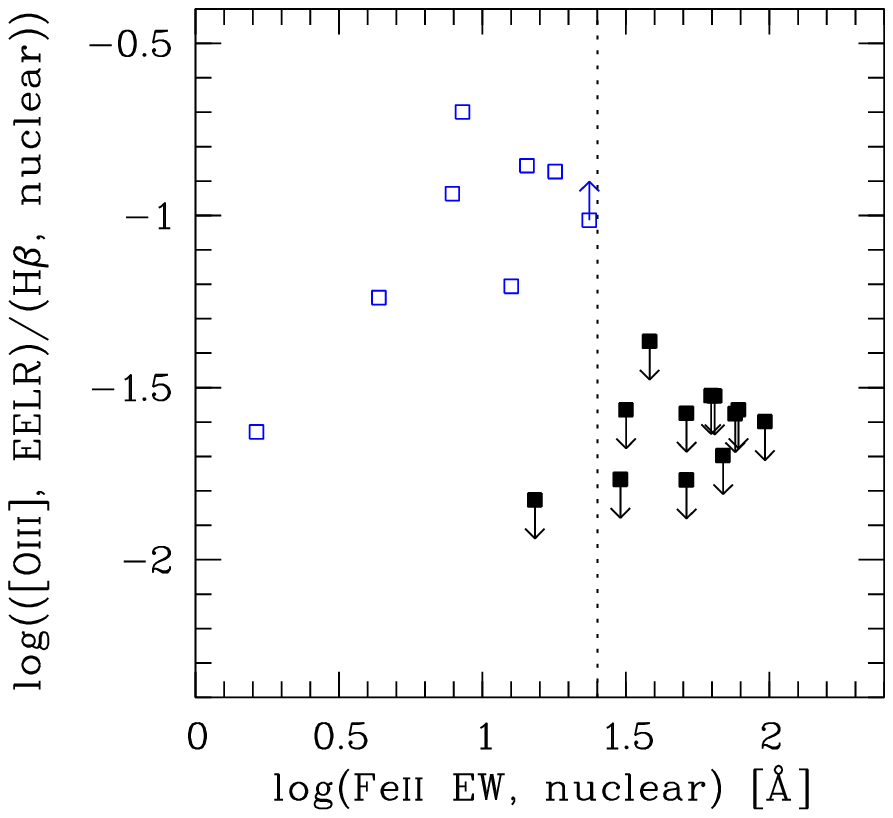}
   \includegraphics[height=4.5cm]{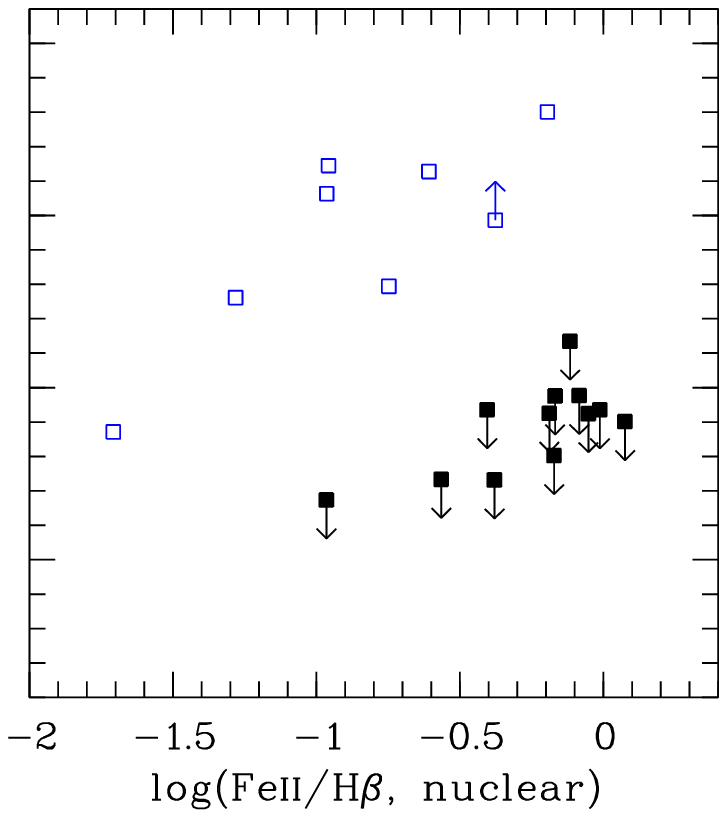}
   }
  \caption{Comparison of the extended \Ox\ flux (again normalised to nuclear
    H$\beta$) with two different measures of \Fe\ strength. Left: Equivalent width
    of the \Fe\ emission; right: \Fe\ flux divided by the broad H$\beta$ flux.
    Symbols as in Fig.~\ref{fig:blr}.}
  \label{fig:blr_fe}
\end{figure}

In essentially all quasars where we detected extended emission in \Ox, we also
see a significant nuclear \Ox\ component. We also measured \Fe\ emission to be
very weak. This is shown in the left panel of Fig.~\ref{fig:blr}, where we
plot the \emph{nuclear} \Ox/\Hb\ flux ratio against the \Fe\ equivalent
width. Apparently, quasars with and without an EELR populate two distinct
regions on this plot. By drawing a vertical line at $\log(\Fe\mathrm{\ EW}) =
1.4$, the two groups are well separated except for a single interloper with
undetected EELR, which might be due to relatively poor observing condition for
that objects (seeing $\sim$1.8\arcsec).  Figure~\ref{fig:blr_fe} compares two
different measures of relative \Fe\ strength with the H$\beta$ normalised flux
ratio.  In the left panel we employ the \Fe\ EW as used already in
Fig.~\ref{fig:blr}, and in the right panel we show the \Fe/H$\beta$
ratio. While the objects with and without detected EELR are again clearly
separated in the former diagram, the distribution of \Fe/H$\beta$ values is
essentially identical for the two groups.

Comparing the normalised EELR flux with the FWHM of the broad H$\beta$
emission line we find an even clearer separation, albeit with a slightly
larger number of interlopers. This is shown in the right-hand panel of
Fig.~\ref{fig:blr}, where 6/8 of all objects with a detected EELR have line
widths well in excess of $\mathrm{FWHM}({\mathrm{H}\beta})\sim 4000$~km/s.
On the other hand, all quasars (except one) with non-detected EELR have
1000~km/s $\la$ $\mathrm{FWHM}({\mathrm{H}\beta})$ $\la$ 3000~km/s.
We note that the one exception, the RQQ \object{PG~2214+139} with
$\mathrm{FWHM}({\mathrm{H}\beta})\sim 5000$\,km/s and no detected EELR, shows
a very peculiar multicomponent H$\beta$ line shape. The two interlopers with
$\mathrm{FWHM}({\mathrm{H}\beta})< 4000$\,km/s have the largest \Fe\ EWs
values among the objects with detected EELR.

\begin{figure}
  \centering
  \resizebox{\hsize}{!}{
    \includegraphics[height=4.5cm]{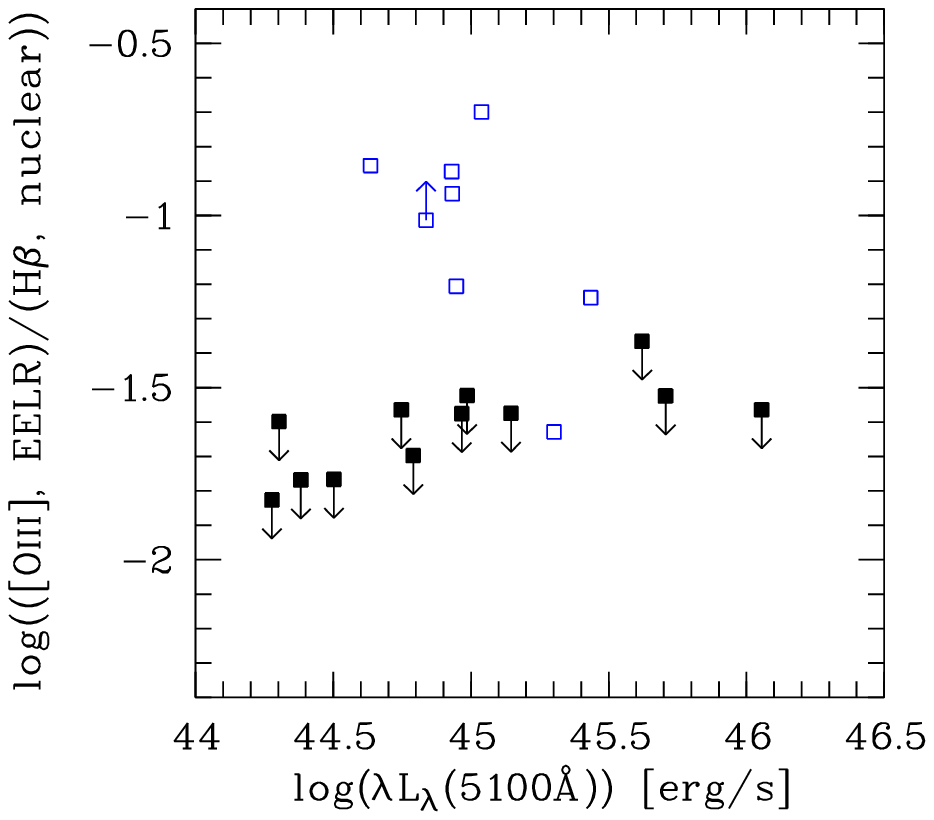}
    \hfill
    \includegraphics[height=4.5cm]{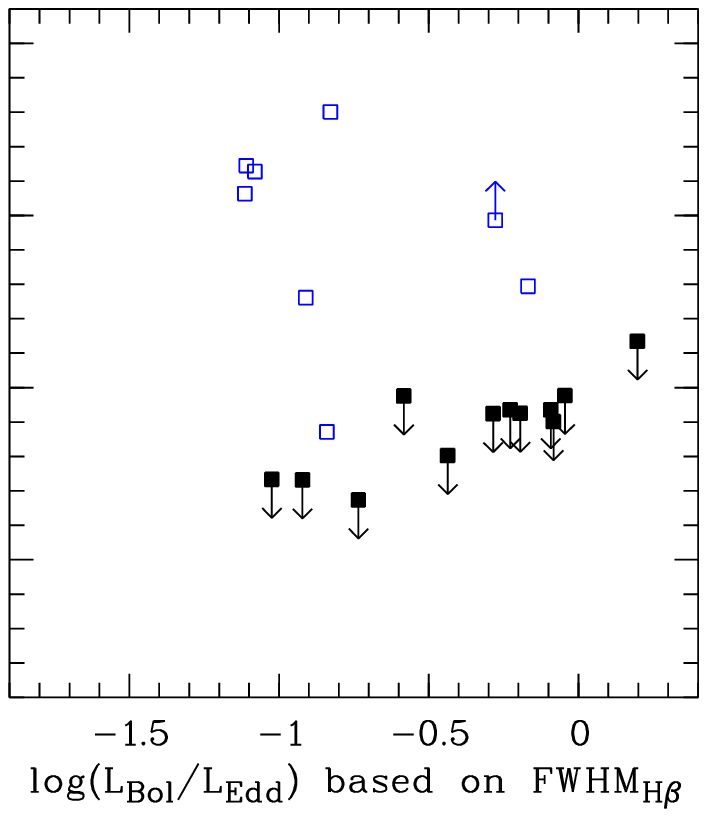}}
  \caption{The EELR flux normalised by the nuclear \Hb\ flux plotted
  against continuum luminosity at 5100\,\AA\ (left), and against Eddington
  ratios (right). Symbols as in Fig.~\ref{fig:blr}.}
  \label{fig:bh}
\end{figure}

\subsection{Luminosities, black hole masses and Eddington ratios}

We now explore whether the above results can be related to more fundamental
quantities such as luminosities, black hole masses and Eddington ratios.
Figure~\ref{fig:bh} presents, in a similar way as in Fig.~\ref{fig:blr_fe},
the EELR flux normalised by the nuclear \Hb\ flux plotted against the
continuum luminosity (left panel) and against the Eddington ratio (right
panel). All objects have broadly similar continuum luminosities, which
basically \textit{rules out} the possibility that it is simply the luminosity
of the quasar determining the EELR properties. The FWHM is then the dominant
term in the computation of $M_\mathrm{BH}$, so that a pattern very similar to
that seen in $\mathrm{FWHM}({\mathrm{H}\beta})$ arises also for
$M_\mathrm{BH}$. We find that a dividing line at
$\log(M_\mathrm{BH}/M_\odot)=8.5$ separates the objects having smaller black
hole masses and non-detected EELRs from the objects with larger masses and
detected EELRs. If on the other hand we consider the Eddington ratios, there is 
again \emph{no} clear separation
between the two groups, with the distribution of
$\log(L_\mathrm{bol}/L_\mathrm{Edd})$ covering the same range for detected and
non-detected QSOs. We reiterate that this fact is independent of our choice of
how to estimate black hole masses (Sect.~\ref{sec:bhmasses}).

\section{Conclusions}

By means of integral field spectroscopy, we have revealed the presence of
extended emission line regions (EELRs) in a sample of predominantly
radio-quiet low-redshift QSOs.  The sizes and luminosities are much higher
than those of the well-known EELRs in Seyfert galaxies, extending out to
several kpc and therefore stretching over the entire host galaxy (and possibly
beyond).  Some 40~\% of the objects in our sample show a prominent EELR, while
we failed to detect any EELR signature in the remaining 60~\%.  
The detection or non-detection of an EELR can be linked to the
\emph{nuclear} spectral properties of the quasar itself, independent of their
radio classification, in excellent agreement with what \citet{Boroson:1984}
and \citet{Boroson:1985} found from their off-nuclear long-slit spectra. 
Our observations go further in that our integral field spectral data map out the
entire nuclear-subtracted EELRs and derive their luminosities.

Correlations between various nuclear spectral properties were already
studied by many authors
\citep[e.g.][]{Boroson:1992,Sulentic:2000,Netzer:2004}.  
Here we provide new evidence that properties of the nuclear spectra,
depending on the sub-pc scale conditions in AGN, are linked to properties of
the host galaxies extending over several kpc. It thus appears possible
to predict with high confidence whether or not a QSO will show a very extended
emission line region just from its nuclear spectrum: If it has broad \Hb\ and
weak \Fe, it will probably have not only strong nuclear, but also strong
extended \Ox\ emission. 

It is tempting to interpret this relation between sub-pc and super-kpc scales
in the context of the current discussion about AGN feedback affecting the host
galaxies. Since we can rule out the QSO luminosity as a main driver of the
differences between presence or absence of an EELR, the remaining 
possibilities are that there could be differences in the structure of the
nuclear region (obscuration in connection with possible inclination effects); 
or differences due to an outflow or jet, 
or the host galaxies of the two QSO classes are intrinsically different, 
in particular with respect to their overall distribution of warm ionised gas. 
We conclude that QSO host galaxies come in at least two
flavours: with or without extended ionized gas. Whether this highlights
different initial conditions or a sequence of evolutionary stages we cannot
say at present. More and better quality data will be needed to elucidate these
intriguing trends.

\begin{acknowledgements}
BH and LW acknowledge financial support by the DFG Priority Program 1177,
grant Wi 1369/22-1.
KJ acknowledges support through the DFG Emmy Noether-Program.
\end{acknowledgements}

\bibliographystyle{aa}
\bibliography{references}
\end{document}